\documentclass[11pt,a4paper,fleqn]{article}

\usepackage{amsmath,amssymb,amsthm,enumerate}

\newcommand{\vspacebefore}{\raisebox{0ex}[2.5ex][0ex]{\null}}
\newcommand{\p}{\partial}
\newcommand{\const}{\mathop{\rm const}\nolimits}
\newcommand{\Equiv}{\mathop{\rm \, equiv}}

\newcommand{\thetbn}{\arabic{nomer}}

\newtheorem{theorem}{Theorem}
\newtheorem{lemma}{Lemma}
\newtheorem{corollary}{Corollary}

{\theoremstyle{definition}

\newtheorem{note}{Note}
\newtheorem*{note*}{Note}
}

\begin{document}

\par\noindent {\LARGE\bf
Group classification\\ of (1+1)-Dimensional
Schr\"odinger Equations\\ with Potentials and Power Nonlinearities
\par}

{\vspace{5mm}\par\noindent {\it
Roman O. POPOVYCH~$^\dag$, Nataliya M. IVANOVA~$^\ddag$ and Homayoon ESHRAGHI~$^\star$
} \par\vspace{2mm}\par}

{\vspace{2mm}\par\noindent {\it
$^{\dag\hspace{-0.2mm},\ddag}$Institute of Mathematics of NAS of Ukraine,
3 Tereshchenkivska Str., 01601 Kyiv-4, Ukraine
} \par}
{\par\noindent {$\phantom{\dag}$~\rm E-mail: }{\it
$^\dag$rop@imath.kiev.ua, $^\ddag$ivanova@imath.kiev.ua
} \par}

{\vspace{2mm}\par\noindent {\it
$^\star$\,\,Institute for Studies in Theor. Physics and Mathematics,
Tehran P.O.\,Box: 19395-5531, Iran
} \par}
{\par\noindent {$\phantom{\dag}$~\rm E-mail: }{\it
eshraghi@theory.ipm.ac.ir
} \par}

{\vspace{6mm}\par\noindent\hspace*{10mm}\parbox{140mm}{\small
We perform the complete group classification
in the class of nonlinear Schr\"odinger equations of the form
$i\psi_t+\psi_{xx}+|\psi|^\gamma\psi+V(t,x)\psi=0$
where $V$ is an arbitrary complex-valued potential
depending on $t$ and $x,$ $\gamma$ is a real non-zero constant.
We construct all the possible inequivalent potentials
for which these equations have non-trivial Lie
symmetries using a combination of algebraic and compatibility methods. The proposed approach can be applied to solving
group classification problems for a number of important classes
of differential equations arising in mathematical physics.
}\par}

\section{Introduction}

Nonlinear Schr\"odinger equations (NSchEs) are important objects for investigation
in different fields of physics and mathematics.
They are used in geometric optics~\cite{Faddeev&Takhtajan1987},
nonlinear quantum mechanics~\cite{Doebner&Goldin1994} and
the theory of Bose--Einstein condensation.
NSchEs also have a number of applications in wave propagation in inhomogeneous medium
and arise as a model of plasma phenomena.
The cubic Schr\"odinger equation is one of the most known integrable models
of mathematical physics.
At the same time the physical interpretation of some known types of nonlinear Schr\"odinger equations
is not completely clear and is an interesting problem to solve.

Schr\"odinger equations have been investigated by means of symmetry methods by a number of authors.
(See e.g.~\cite{Niederer1972,Gagnon93,Niederer1973,Miller1977,Fushchych&Moskaliuk1981,Gagnon88,
Gagnon89a,Gagnon89b,Gagnon89c,Ivanova2002} and references therein
for classical Lie symmetries.)
In fact, group classification for Schr\"odinger equations was first performed
by S.~Lie. More precisely, his classification~\cite{Lie1881withTrans}
of all the linear equations
with two independent complex variables contains, in an implicit form,
solving the classification problem for the linear (1+1)-dimensional
Schr\"odinger equations with an arbitrary potential. And it is follows from Lie's proof
that the equations for the harmonic and repulsive oscillators and free fall are
locally equivalent to the free Schr\"odinger equation.

To the best of our knowledge, actual investigations of Lie symmetries for
Schr\"odinger equations were started in 1970s
with the linear case~\cite{Niederer1972,Niederer1973,Miller1977}.
The next considered class covered $(1+n)$-dimensional NSchEs with nonlinearities of
the form $f(|\psi|)\psi,$ which are notable for their symmetry properties
because any such equation is invariant with respect to the Galilean group.
It turned out that extensions of this invariance group are possible only for
the logarithm and power functions, and there exists
the power value $\gamma=4/n$
which is special with respect to the symmetry point of view~\cite{Fushchych&Moskaliuk1981}.
Namely, the free Schr\"odinger equation and
the NSchE with the nonlinearity~$|\psi|^{4/n}\psi$ are distinctive ones from a lot of
similar equations, since they admit
the complete Galilei group extended with both the scale and conformal transformations.
(Here $n$ is the number of spatial variables,
and for $n=1$ and $n=2$ they are the quintic and cubic equations respectively that stand out
against the other NSchEs.) This NSchE has also other special properties,
and the value $\gamma=4/n$ is called now the critical power.

The results mentioned above formed a basis for symmetry studying
more extended classes of NSchEs.
So, finishing the series of papers~\cite{Gagnon88,Gagnon89a,Gagnon89b,Gagnon89c,Gagnon93}
on group analysis and exact solutions of NSchEs,
L.~Gagnon and P.~Winternitz~\cite{Gagnon93} investigated a general class
of (1+1)-dimensional variable coefficient cubic SchEs.
It is the symmetry approach that was applied by H.-D.~Doebner and G.A.~Goldin
to obtain new equations which generalize the Schr\"odinger equation
and can be used in nonlinear quantum mechanics~\cite{Doebner&Goldin1994}.
These equations were investigated in more detail with the symmetry point of view
by a number of authors~\cite{Nattermann&Doebner1996,Doebner&Goldin&Nattermann1999,Zhdanov&Roman}.
The complete group classification of constant coefficient NSchEs with nonlinearities
of the general form $F=F(\psi,\psi^*)$ was performed
by A.G.~Nikitin and R.O.~Popovych~\cite{Nikitin&Popovych}.

Group classification is one of symmetry methods used to choose physically relevant models
from parametric classes of systems of (partial or ordinary) differential equations.
The parameters can be constants or functions of independent variables, unknown functions and their derivatives.
Exhaustive consideration of the problem of group classification for a parametric class $\mathcal{L}$
of systems of differential equations includes the following steps:
\begin{enumerate}
\item
Finding the group $G^{\ker}$ (the kernel of maximal invariance groups)
of local transformations that are symmetries for all systems from $\mathcal{L}$.
\item
Construction of the group $G^{\Equiv}$ (the equivalence group) of local transformations
which transform $\mathcal{L}$ into itself.
\item
Description of all possible $G^{\Equiv}$-inequivalent values of parameters that admit
maximal invariance groups wider than $G^{\ker}$.
\end{enumerate}
Following S.~Lie, one usually considers infinitesimal transformations instead of finite ones.
This approach essentially simplifies the problem of group classification,
reducing it to problems for Lie algebras of vector fields.
See~\cite{Zhdanov&Roman,Nikitin&Popovych,Zhdanov&Lahno1999,Ovsiannikov1982,%
Popovych&Yehorchenko2001a,Popovych&Ivanova2003a,Akhatov&Gazizov&Ibragimov1987,Akhatov&Gazizov&Ibragimov1989}
for precise formulation of group classification problems
and more details on the used methods.

In this paper we study a class of NSchEs of the form
\begin{equation}\label{schr}
i\psi_t+\psi_{xx}+|\psi|^\gamma\psi+V\psi=0,
\end{equation}
where the potential $V=V(t,x)$ is an arbitrary complex-valued
smooth function of the variables~$t$ and~$x,$
$\gamma$ is a real non-zero constant.
(Here and below subscripts of functions denote differentiation
with respect to the respective variables.)
To find a complete set of inequivalent cases of $V$
admitting extensions of the maximal Lie invariance algebra,
we combine the classical Lie approach, studying the algebra
generated by all the possible Lie symmetry operators for equations
from class~(\ref{schr}) (the adjoint representation, the inequivalent
one-dimensional subalgebras etc.)
and investigation of compatibility of classifying equations.
The subclass of~(\ref{schr}) where $\gamma=2$ (the cubic SchEs with potentials)
has been investigated in~\cite{Popovych&Ivanova&Eshraghi2003CubicLanl}
in a similar way.

In fact, here we solve three classification problems for different classes of
equations having the form~(\ref{schr}):
with the potentials depending only on $t$ (Section~III),
with the stationary potentials (Section~IV) and
the general case with arbitrary potentials (Section~II).
Moreover, it is proved in Section~II the constant $\gamma$ can be assumed as \emph{fixed}
under our consideration.
And there exists a special constant~$\widehat\gamma$
depending on the power~$\gamma$ ($\widehat\gamma=\gamma^{-1}(4-\gamma)$),
which arises at the beginning of the classification procedure
when classification condition~(\ref{classcondforcshewp}) is constructed
and explicitly appears in the final results of classification.
The value $\gamma=4$ (quintic nonlinearity, it is the same that $\widehat\gamma=0$)
is special with respect to group classification in class~(\ref{schr}).

The classification approach used in this paper allows us to formulate
a necessary and sufficient condition of mutual equivalence for the cases of
extensions of maximal invariance algebras
in  algebraic terms (Corollary~\ref{criteria.equiv.cshewp}).
The classical stationary potentials (free particle, the harmonic and repulsive
oscillators, free fall,  radial free particle, the radial harmonic and repulsive
oscillators~\cite{Miller1977}) naturally arise under the group classification
with respect to the (smaller) equivalence group in the class of stationary potentials.
Using Corollary~\ref{criteria.equiv.cshewp} and
the complete equivalence group of class~(\ref{schr})
we easily construct transformations of these $x$-dependent potentials to
$x$-free ones in explicit form (see Remark~\ref{ExplicitFormForTransVtVx}).

\section{General Case {\mathversion{bold}$V=V(t,x)$}}

Consider an operator
$Q=\xi^t\partial_t+\xi^x\partial_x+\eta\partial_\psi+\eta^*\partial_{\psi^*}$
from the maximal Lie invariance algebra $A^{\max}(\gamma,V)$ of equation~(\ref{schr})
with a power $\gamma$ and a potential $V.$
Here $\xi^t,$ $\xi^x$ and $\eta$ are smooth functions of $t,$ $x,$ $\psi$ and $\psi^*.$
The infinitesimal invariance condition~\cite{Ovsiannikov1982,Olver1982}
of equation~(\ref{schr}) with respect to the operator~$Q$
implies the linear overdetermined system on the coefficients of~$Q$:
\begin{gather*}
\xi^t_\psi=\xi^t_{\psi^*}=\xi^t_x=0,\quad
\xi^x_\psi=\xi^x_{\psi^*}=0,\quad
\xi^t_t=2\xi^x_x,\quad
\eta_{\psi^*}=\eta_{\psi\psi}=0, \quad \psi\eta_\psi=\eta,\\
2\eta_{\psi x}=i\xi^x_t,\quad
\gamma(\eta_\psi+\eta^*_\psi)=-2\xi^t_t,\quad
i\eta_{\psi t}+\eta_{\psi xx}+\xi^tV_t+\xi^xV_x+\xi^t_tV=0.
\end{gather*}

Therefore, the following theorem holds.

\begin{theorem}
Any operator~$Q$ from $A^{\max}(\gamma,V)$
of equation~(\ref{schr}) with arbitrary potential $V$
lies in the linear span of operators of the form
\begin{equation}\label{gflsopscshewp}
D(\xi)=\xi\p_t+\dfrac12\,\xi_tx\p_x+\dfrac18\,\xi_{tt}x^2M-\dfrac1{\gamma}\,\xi_tI,\quad
G(\chi)=\chi\p_x+\dfrac12\chi_txM,\quad
\lambda M.
\end{equation}
Here
$\chi=\chi(t),$ $\xi=\xi(t)$ and $\lambda=\lambda(t)$
are arbitrary smooth functions of $t,$
$M=i(\psi\p_{\psi}-\psi^*\p_{\psi^*})$,
$I=\psi\p_{\psi}+\psi^*\p_{\psi^*}.$
Moreover, the coefficients of $Q=D(\xi)+G(\chi)+\lambda M\in A^{\max}(\gamma,V)$
should satisfy the classifying condition
\begin{equation}\label{classcondforcshewp}
\xi V_t+\left(\frac12\xi_tx+\chi\right)V_x+\xi_tV=
\frac18\,\xi_{ttt}x^2+\frac12\chi_{tt}x+\lambda_t+i\frac{\widehat\gamma}4\,\xi_{tt}.
\end{equation}
Here and below $\widehat\gamma=\gamma^{-1}(4-\gamma).$
\end{theorem}

\begin{theorem}
The Lie algebra of the kernel of maximal Lie invariance groups of equations
from class~(\ref{schr}) is $A^{\mathrm {ker}}=\langle M \rangle.$
\end{theorem}

\begin{note}
Sometimes (e.g. for reduction and construction of solutions)
it is convenient to use the amplitude~$\rho$ and the phase~$\varphi$
instead of the wave function~$\psi=\rho e^{i \varphi}.$ Then equation~(\ref{schr})
is replaced by the system for two real-valued functions~$\rho$ and~$\varphi$:
\[
\rho_t+2\rho_x\varphi_x+\rho\varphi_{xx}+\rho\mathop{\rm Im}\nolimits V=0,\qquad
-\rho\varphi_t-\rho(\varphi_x)^2+\rho_{xx}+\rho^{\gamma+1}+\rho\mathop{\rm Re}\nolimits V=0.
\]
Operators~(\ref{gflsopscshewp}) have the same form with
$M=\partial_\varphi,$ $I=\rho\partial_\rho.$
\end{note}

To study equivalence transformations for class~(\ref{schr}),
both the infinitesimal and direct method are used.
In the framework of the infinitesimal method we consider
a first-order differential operator of the most general form
in the space of the variables
$t,$ $x,$ $\psi,$ $\psi^*,$ $V,$ $V^*$ and $\gamma,$ i.e.
\[
Q=\xi^t\partial_t+\xi^x\partial_x+\eta\partial_\psi+\eta^*\partial_{\psi^*}
+\theta\partial_V+\theta^*\partial_{V^*}+\Gamma\partial_\gamma,
\]
where $\xi^t,$ $\xi^x,$ $\eta,$ $\theta$ and $\Gamma$ may depend on
all the variables, and assume it being an infinitesimal symmetry operator for
the system
\begin{equation}\label{NSchEwithPNPsystemForEquivTrans}
i\psi_t+\psi_{xx}+|\psi|^\gamma\psi+V\psi=0,\quad
\gamma_t=\gamma_x=\gamma_\psi=\gamma_{\psi^*}=0,\quad
V_\psi=V_{\psi^*}=0.
\end{equation}
(Under the prolongation procedure for equivalence transformations we suppose
$\psi$ is a function of $t$ and $x$ as well as
$V$ and $\gamma$ are functions of $t,$ $x$ and $\psi.$)

\begin{theorem}
The Lie algebra~$A^{\Equiv}$ of the equivalence group~$G^{\Equiv}$ of class~(\ref{schr})
is generated by the operators
\[\arraycolsep=0ex\begin{array}{l}\displaystyle
D'(\xi)=D(\xi)+\dfrac18\,\xi_{ttt}x^2(\partial_V+\partial_{V^*})
+\dfrac i{\gamma}\,\xi_{tt}(\partial_V-\partial_{V^*})-
\xi_t(V\partial_V+{V^*}\partial_{V^*}),\\[2ex]
G'(\chi)=G(\chi)+\dfrac12\chi_{tt}x(\partial_V+\partial_{V^*}),\quad
M'(\lambda)=\lambda M+\lambda_t(\partial_V+\partial_{V^*}).
\end{array}\]
\end{theorem}

In the framework of the direct method we look for all local
transformations in the space of the variables
$t,$ $x,$ $\psi,$ $\psi^*,$ $V,$ $V^*$ and $\gamma,$
which preserve system~(\ref{NSchEwithPNPsystemForEquivTrans}).

\begin{theorem}\label{NSchEwithPNPTheoremOnGequiv}
The equivalence group~$G^{\Equiv}$ of the class~(\ref{schr}) is generated by
the family of continuous transformations
\begin{equation}\label{eqtranspcshe}
\arraycolsep=0ex\begin{array}{l}\displaystyle
\tilde t=T, \quad
\tilde x=x\sqrt{T_t}+X,
\quad
\tilde \psi=\psi\dfrac{1}{\sqrt{T_t}}
\exp\left(\dfrac i8\dfrac{T_{tt}}{T_t}\,x^2+
\dfrac i2\dfrac{X_{t}}{\sqrt{T_t}}\,x +i\Psi \right)\!, \quad
\tilde\gamma=\gamma,
\\[3ex]
\tilde V=\dfrac1{T_t}\left(V+\dfrac 18\left(\dfrac{T_{tt}}{T_t}\right)_{\!\!t}x^2
+\dfrac{1}2\left(\dfrac{X_{t}}{\sqrt{T_t}}\right)_{\!\!t}x
+i\dfrac{\widehat\gamma}4\,\dfrac{T_{tt}}{T_t}
-\left(\dfrac 14\dfrac{T_{tt}}{T_t}\,x+
\dfrac {1}2\dfrac{X_{t}}{\sqrt{T_t}}\right)^2
\!+\Psi_t\right)
\end{array}\end{equation}
and two discrete transformations:
the space reflection $I_x$
($\tilde t=t,$ $\tilde x=-x,$ $\tilde \psi=\psi,$ $\tilde\gamma=\gamma,$ $\tilde V=V$)
and the Wigner time reflection $I_t$
($\tilde t=-t,$ $\tilde x=x,$ $\tilde \psi=\psi^*,$ $\tilde\gamma=\gamma,$ $\tilde V=V^*$).
Here $T$, $X$ and $\Psi$ are arbitrary smooth functions of $t,$ $T_t>0.$
\end{theorem}

We also prove the stronger statement than Theorem~\ref{NSchEwithPNPTheoremOnGequiv}.

\begin{theorem}\label{NSchEwithPNPTheoremOnCompleteEquiv}
If two equations from class~(\ref{schr}) with the parameter values $(\gamma, V)$ and $(\tilde\gamma,\tilde V)$
are transformed each to other by local transformations then $\tilde\gamma=\gamma$.
Moreover, since $\gamma\ne0$ any transformation of such type belongs to~$G^{\Equiv}$.
\end{theorem}

\begin{note}
It follows from Theorems~\ref{NSchEwithPNPTheoremOnGequiv} and~\ref{NSchEwithPNPTheoremOnCompleteEquiv}
that there exist no equivalence and, moreover, local transformations changing $\gamma.$
Therefore, we can assume that \emph{$\gamma$ is fixed} in our consideration below
and omit it from notations of the maximal Lie invariance algebras of an equation
of form~(\ref{schr}) etc.
\end{note}

\begin{note}
The linear span of operators of the form~(\ref{gflsopscshewp}) ($\gamma$ is fixed!)
is an (infinite-dimensional) Lie algebra~$A^{\cup}$ under the usual Lie bracket
of vector fields. Since for any $Q\in A^{\cup}$ where $(\xi^t,\xi^x)\ne(0,0)$
we can find $V$ satisfying condition~(\ref{classcondforcshewp}) then
$A^{\cup}=\langle\,\bigcup_V A^{\max}(V)\,\rangle.$
The non-zero commutation relations between the basis elements of~$A^{\cup}$
are the following ones:
\[\arraycolsep=0ex\begin{array}{l}\displaystyle
[D(\xi^1),D(\xi^2)]=D(\xi^1\xi^2_t-\xi^2\xi^1_t),\quad
[D(\xi),G(\chi)]=G\left(\xi\chi_t-\frac12\xi_t\chi\right),\quad
[D(\xi),\lambda M]=\xi\lambda_t M,\\[1ex]\displaystyle
[G(\chi^1),G(\chi^1)]=\frac12(\chi^1\chi^2_t-\chi^2\chi^1_t)M.
\end{array}\]
We use the notation $\mathop{\rm Aut}\nolimits (A^{\cup})$ for
the automorphism group acting on $A^{\cup},$
which is generated by all the one-parameter groups corresponding to
the adjoint representations of operators of~$A^{\cup}$ into~$A^{\cup}$
and two discrete transformations $\mathop{\rm Ad}\nolimits I_x$
and $\mathop{\rm Ad}\nolimits I_t$ included additionally.
The actions of $\mathop{\rm Ad}\nolimits I_x$ and $\mathop{\rm Ad}\nolimits I_t$
on the basis elements of $A^{\cup}$ are defined by the formulas
$\mathop{\rm Ad}\nolimits I_x\ G(\chi)=G(-\chi)$
(the other basis operators do not change) and
$\mathop{\rm Ad}\nolimits I_t\ D(\xi)=D(\tilde\xi)$,
$\mathop{\rm Ad}\nolimits I_t\ G(\chi)=G(\tilde\chi),$
$\mathop{\rm Ad}\nolimits I_t\ \lambda M=\tilde\lambda M$,
where $\tilde\xi(t)=-\xi(-t)$, $\tilde\chi(t)=\chi(-t)$ and $\tilde\lambda(t)=-\lambda(-t)$.
\end{note}

\begin{corollary}
$A^{\Equiv}\simeq A^{\cup},$ $G^{\Equiv}\simeq\mathop{\rm Aut}\nolimits (A^{\cup}),$
and the isomorphism is determined by means of prolongation of operators
from $A^{\cup}$ to the space~$(\gamma,V,V^*).$

\end{corollary}

\begin{corollary}\label{criteria.equiv.cshewp}
Let $A^1$ and $A^2$ be the maximal Lie invariance algebras of
equations from class~(\ref{schr}) for some potentials,
and ${\cal V}^i=\{\,V\,|\,A^{\max}(V)=A^i\},$ $i=1,2.$
Then ${\cal V}^1\sim {\cal V}^2\!\!\mod\!G^{\Equiv}$ iff
$\,A^1\sim A^2\!\!\mod\!\mathop{\rm Aut}\nolimits (A^{\cup}).$
\end{corollary}

\begin{lemma}\label{subalgsAcup}
A complete list of $\mathop{\rm Aut}\nolimits A^{\cup}$-inequivalent one-dimensional subalgebras
of~$A^{\cup}$ is exhausted by the algebras
$\langle\partial_t\rangle,$ $\langle\partial_x\rangle,$
$\langle tM\rangle,$ $\langle M\rangle.$
\end{lemma}

\noindent {\it Proof:}
Consider any operator $Q\in A^{\cup},$ i.e. $Q=D(\xi)+G(\chi)+\lambda M.$
Depending on the values of~$\xi,$ $\chi$ and~$\lambda$ it is equivalent
under~$\mathop{\rm Aut}\nolimits (A^{\cup})$ and multiplication by a number
to one from the following operators:
$D(1)$ if $\xi\ne0;\;$
$G(1)$ if $\xi=0$ and $\chi\ne0;\;$
$tM$ if $\xi=\chi=0,$ $\lambda_t\ne0;\;$
$M$ if $\xi=\chi=\lambda_t=0.$

\begin{corollary}\label{lemma.vtvx0}
If $\,A^{\max}(V)\ne A^{\ker}\,$ then $\,V_tV_x=0\!\!\mod G^{\Equiv}$.
\end{corollary}

\noindent {\it Proof:}
Under the corollary assumption there exists
an operator $Q=D(\xi)+G(\chi)+\lambda M\in A^{\max}(V)$
which do not belong to $\langle M\rangle.$
Condition~(\ref{classcondforcshewp}) implies $(\xi,\chi)\ne(0,0).$
Therefore, in force of Lemma~\ref{subalgsAcup}
$\;\langle Q\rangle\sim \langle\partial_t\rangle$ or
$\langle\partial_x\rangle\!\!\mod \mathop{\rm Aut}\nolimits A^{\cup},$
i.e. $\,V_tV_x=0\!\!\mod G^{\Equiv}$.

\begin{theorem}\label{theorem.gc.pcshe}
A complete set of inequivalent cases of $\,V$
admitting extensions of the maximal Lie invariance algebra
of equations~(\ref{schr}) is exhausted by the potentials given in Table~1.
\end{theorem}

\newcounter{tbn} \setcounter{tbn}{0}
{\begin{center}
Table 1. Results of classification.
Here $W(t),\nu,\alpha,\beta\in\mathbb{R},$ $(\alpha,\beta)\ne(0,0).$
\\[1.5ex] \footnotesize
\setcounter{tbn}{0}
\renewcommand{\arraystretch}{1.4}
\begin{tabular}{|r|c|l|}
\hline\vspacebefore
N &$V$ &\hfill {Basis of $A^{\max}$\hfill} \\
\hline\vspacebefore
\thetbn& $V(t,x)$ & $M\;$ \\
\hline\vspacebefore
\refstepcounter{tbn}\thetbn\label{pcsheV1}& $iW(t)$ & $M,\;$ $\p_x,\;$ $G(t)$\\[0.8ex]
\refstepcounter{tbn}\thetbn\label{pcsheV2}&
$\dfrac i2\dfrac{\widehat\gamma\, t+\nu}{t^2+1}$
& $M,\;$ $\p_x,\;$ $G(t),\;$ $D(t^2+1)$\\[1.3ex]
\refstepcounter{tbn}\thetbn\label{pcsheV3}& $i\nu t^{-1}\!,$\quad
$\nu\ne0,\dfrac{\widehat\gamma}{2}$
& $M,\;$ $\p_x,\;$ $G(t),\;$ $D(t)$\\
\refstepcounter{tbn}\thetbn\label{pcsheV4}& $i$
& $M,\;$ $\p_x,\;$ $G(t),\;$ $\p_t$ \\
\refstepcounter{tbn}\thetbn\label{pcsheV5}& $0,\:$ $\gamma\ne4$
& $M,\;$ $\p_x,\;$ $G(t),\;$ $\p_t,\;$ $D(t)$\\
& $\phantom{0,\:}$ $\gamma=4$
& $M,\;$ $\p_x,\;$ $G(t),\;$ $\p_t,\;$ $D(t),\;$ $D(t^2)$\\
\refstepcounter{tbn}\thetbn\label{pcsheV6}& $V(x)$
& $M,\;$ $\p_t$\\
\refstepcounter{tbn}\label{pcsheV7}\thetbn&
$(\alpha+i\beta)x^{-2},\:$ $\gamma\ne4$ & $M,\;$ $\p_t,\;$ $D(t)$\\
&$\phantom{(\alpha+i\beta)x^{-2},\:}$ $\gamma=4$ & $M,\;$ $\p_t,\;$ $D(t),\;$ $D(t^2)$\\
\hline
\end{tabular}
\end{center}}

For convenience we use below the double numeration T.N of classification cases
where T is a table number and N is a row number.

\begin{note}
We mean that the invariance algebras for Cases~1.0, 1.\ref{pcsheV1}, 1.\ref{pcsheV6} and
analogous ones from Tables~2 and~3 are maximal if these cases are inequivalent under the
corresponding equivalence group to the other, more specialized, cases
from the same table.
\end{note}

\begin{note}
There exists a discrete equivalence transformation $\tau$
for the set of potentials $i\nu t^{-1},$ $\nu\in\mathbb{R},$
which has form~(\ref{eqtranspcshe})
with $T=-t^{-1},$ $X=0,$ $\Psi=0.$
It transforms $\nu$ in the following way: $\nu\to\widehat\gamma/2-\nu.$
For the cases under consideration to be completely inequivalent,
we have  to assume additionally that $\nu\ge\widehat\gamma/4$
(or $\nu\le\widehat\gamma/4$) in Case~1.3.
Since $I_t\in G^{\Equiv}$ we can assume analogously $\nu\ge0$ in Case~1.2 and $\beta\ge0$ in Case~1.7.
Moreover, $\tau$ is a discrete symmetry transformation
for Case~1.\ref{pcsheV3} ($\nu=\widehat\gamma/4$) and,
as a limit of the continuous transformations generated by the operator $D(t^2+1),$
for Case~1.\ref{pcsheV2}.
\end{note}

If we use Corollary~\ref{lemma.vtvx0} then to prove Theorem~\ref{theorem.gc.pcshe} it
is sufficient to study two cases: $V_x=0$ and $V_t=0$.
In fact, below we obtain the complete results of group classifications
for both special cases and then unite them for the general case under consideration.

\section{Case {\mathversion{bold}$V=V(t)$}}

Consider the equations from class~(\ref{schr}) with potentials satisfying
the additional assumption $V_x=0,$ i.e. $V=V(t).$
The following chain of lemmas gives complete solving of classification problem
in this subclass.

\begin{lemma}
$A^{\mathrm {ker}}_{V_x^{\rule{0mm}{1.7mm}}=0} = \langle M,G(1),G(t)\rangle$.
\end{lemma}
\begin{lemma}
$A^{\Equiv}_{V_x^{\rule{0mm}{1.5mm}}=0} =
\langle M'(\lambda)\, \forall\lambda=\lambda(t),G'(1),G'(t),D'(1),D'(t),D'(t^2)\rangle$.
$G^{\Equiv}_{V_x^{\rule{0mm}{1.5mm}}=0}$\vspace{0.3ex}
is generated by $I_t$, $I_x$ and the transformations of form~(\ref{eqtranspcshe}), where
$T=(a_1t+a_0)/(b_1t+b_0),$ $X=c_1t+c_0,$
$\Psi$ is an arbitrary smooth function of $t$.
$a_i,$ $b_i$ and $c_i$ are arbitrary constants such that $\,a_1b_0-b_1a_0>0.$
\end{lemma}
\begin{lemma}
For any $V=V(t)$:
$V\sim iW\!\!\mod G^{\Equiv}_{V_x^{\rule{0mm}{1.5mm}}=0}$ where $W=\mathrm{Im}V,$
i.e. $W=W(t)\in\mathbb{R}.$
\end{lemma}
\begin{lemma}
$A^{\mathrm {ker}}_{\{iW\}^{\rule{0mm}{1.5mm}}}\! =
A^{\mathrm {ker}}_{V_x^{\rule{0mm}{1.5mm}}=0}.$
$\:A^{\max}(iW)\subset
A^{\cup}_{\{iW\}^{\rule{0mm}{1.5mm}}}\!=A^{\mathrm {ker}}_{\{iW\}^{\rule{0mm}{1.5mm}}}
\:\mbox{$\supset$\hspace{-1.9ex}$+$}\:S\:$
where $\:S=\langle D(1), D(t), D(t^2)\rangle.$
$A^{\cup}_{\{iW\}^{\rule{0mm}{1.5mm}}}\! =
\bigcup_W A^{\max}(iW).$
\vspace{0.4ex}
$A^{\Equiv}_{\{iW\}^{\rule{0mm}{1.5mm}}}\! =
\langle M,G'(1),G'(t),D'(1),D'(t),D'(t^2)\rangle$.
$\left.G^{\Equiv}_{\{iW\}^{\rule{0mm}{1.5mm}}}\!=
G^{\Equiv}_{V_x^{\rule{0mm}{1.5mm}}=0}\right|_{\Psi=\mathrm{const}}\!\!.$
$A^{\cup}_{\{iW\}^{\rule{0mm}{1.5mm}}}\!\simeq
A^{\Equiv}_{\{iW\}^{\rule{0mm}{1.5mm}}}=
\mathop{\mathrm{pr}}_{(V,V^*)}A^{\cup}_{\{iW\}^{\rule{0mm}{1.5mm}}}.$
\end{lemma}
\begin{lemma}
$S\simeq sl(2,\mathbb{R}).$
The complete list of $A^{\cup}_{\{iW\}^{\rule{0mm}{1.5mm}}}\!$-inequivalent
proper subalgebras of $S$ is exhausted by the algebras
$\langle D(1)\rangle,$ $\langle D(t)\rangle,$ $\langle D(t^2+1)\rangle,$
$\langle D(1), D(t)\rangle.$
\end{lemma}
\begin{lemma}\label{criteria.equiv.cshewpdot}
Let $A^1$ and $A^2$ be the maximal Lie invariance algebras of
equations from class~(\ref{schr}) for some potentials from $\{iW(t)\}$,
and ${\cal W}^i=\{\,W(t)\,|\,A^{\max}(iW)=A^i\},$ $i=1,2.$
Then ${\cal W}^1\sim {\cal W}^2\!\!\mod\!G^{\Equiv}_{\{iW\}^{\rule{0mm}{1.5mm}}}$
iff
$\,A^1\bigcap S\sim A^2\bigcap S\!\!\mod\!\mathop{\rm Aut}\nolimits (S).$
\end{lemma}
\begin{lemma}
If $A^{\max}_{\{iW\}^{\rule{0mm}{1.5mm}}}\! \ne
A^{\mathrm {ker}}_{V_x^{\rule{0mm}{1.5mm}}=0}$ the potential $iW(t)$
is $G^{\Equiv}_{\{iW\}^{\rule{0mm}{1.5mm}}}\!$-equivalent
to one from Cases~\mbox{1.2--1.5}.
\end{lemma}

\begin{note}
If $\gamma\ne4$ or $W\ne\const$ $A^{\max}(iW)\not\supset S$ (otherwise,
condition~(\ref{classcondforcshewp}) would imply an incompatible system for $W$).
If $W\!=\!\mathrm{const}$ $\,W\!\in\!\{0,1\}\!\!\mod
G^{\Equiv}_{\{iW\}^{\rule{0mm}{1.5mm}}}$
(Cases~1.5 and~1.4 correspondingly).
Cases~1.2$\nu$ and~1.2$\tilde\nu$ (1.3$\nu$ and~1.3$\tilde\nu$ where $\nu,\tilde\nu\ge\frac14$)
are $G^{\Equiv}$-inequivalent if $\nu\ne\tilde\nu.$
Since $D(t^2+1)$ cannot be contained in any two-dimensional subalgebra of $S$
it is not possible to extend $A^{\max}$ in Case~1.2.
There are two possibilities for extension of $A^{\max}(i\nu t^{-1})$, namely with either
$D(1)$ (for $\nu=0$, Case~1.5) or $D(t^2)$ (for $\nu=(4-\gamma)/(2\gamma)$
that is equivalent to Case~1.5 with respect to $G^{\Equiv}_{\{iW\}^{\rule{0mm}{1.5mm}}}$).
That is why for $\nu=0,$ $\gamma=4$ dimension of $A^{\max}$ is greatest.
\end{note}

\section{Case {\mathversion{bold}$V=V(x)$}}

Consider class~(\ref{schr}) with the additional assumption $V_t=0,$ i.e. $V=V(x).$

\begin{lemma}
$A^{\mathrm {ker}}_{V_t^{\rule{0mm}{1.7mm}}=0} = \langle M,D(1)\rangle$.
\end{lemma}
\begin{lemma}
$A^{\Equiv}_{V_t^{\rule{0mm}{1.5mm}}=0} =
\langle M'(1), M'(t),G'(1),D'(1),D'(t)\rangle$.
$G^{\Equiv}_{V_t^{\rule{0mm}{1.5mm}}=0}$
is generated by  $I_t$, $I_x$ and the transformations of form~(\ref{eqtranspcshe})  where
$T_{tt}=X_t=\Psi_{tt}=0.$
\end{lemma}
\begin{lemma}\label{classification.cshewp.Vt.0}
If $A^{\max}(V) \ne
A^{\mathrm {ker}}_{V_t^{\rule{0mm}{1.5mm}}=0}$ the potential $V(x)$
is $G^{\Equiv}_{V_t^{\rule{0mm}{1.5mm}}=0}\!$-equivalent
to one from cases of either Table~2 if $\gamma\ne4$ or Table~3 if $\gamma=4$.
(Since $I_t\in G^{\Equiv}$ we can assume $\nu\ge0$ in Cases~2.5, 2.6, 3.5--3.7,
$\nu>0$ in Case~2.4
and $\beta\ge0$ in Cases~2.1, 3.1--3.3.)
\end{lemma}

\vspace{-1ex}
{\begin{center}
Table 2. Classification of the subclass $V=V(x)$ if $\gamma\ne4.$\\
Here $\nu,\alpha,\beta\in\mathbb{R},$ $(\alpha,\beta)\ne(0,0).$
\\[1ex]\footnotesize

\renewcommand{\arraystretch}{1.3}\setcounter{tbn}{0}
\begin{tabular}{|c|c|c|l|}
\hline\vspacebefore
N & N$_1\!\!$& $V$ & \hfill {Basis of $A^{\max}$\hfill} \\
\hline
\thetbn\vspacebefore& \ref{pcsheV6} &
$V(x)$ &  $M,\;$ $\p_t$ \\
\hline
\refstepcounter{tbn}\label{pcsheVx1}\thetbn& \ref{pcsheV7} &\vspacebefore
$(\alpha+i\beta)x^{-2}$ &  $M,\;$ $\p_t,\;$ $D(t)$\\
\refstepcounter{tbn}\label{pcsheVx2}\thetbn& \ref{pcsheV7} &
$x^2+i\widehat\gamma+(\alpha+i\beta)x^{-2}$
&   $M,\;$ $\p_t,\;$ $D(e^{4t})$\\
\refstepcounter{tbn}\label{pcsheVx3}\thetbn& \ref{pcsheV4} &
$i$ &  $M,\;$ $\p_t,\;$ $\p_x,\;$ $G(t)$ \\
\refstepcounter{tbn}\label{pcsheVx4}\thetbn& \ref{pcsheV4} &
$x+i\nu,\;$ $\nu\ne 0$ &  $M,\;$ $\p_t,\;$ $\p_x+tM,\;$ $G(2t)+t^2M$\\
\refstepcounter{tbn}\label{pcsheVx5}\thetbn& \ref{pcsheV2} &
$-x^2+i\nu$ &  $M,\;$ $\p_t,\;$ $G(\sin2t),\;$ $G(\cos2t)$\\
\refstepcounter{tbn}\label{pcsheVx6}\thetbn& \ref{pcsheV3} &
$x^2+i\nu,\;$  $\nu\ne\pm\widehat\gamma$ &  $M,\;$ $\p_t,\;$ $G(e^{2t}),\;$ $G(e^{-2t})$\\
\refstepcounter{tbn}\label{pcsheVx7}\thetbn& \ref{pcsheV5} &
0 &  $M,\;$ $\p_t,\;$ $\p_x,\;$ $G(t),\;$ $D(t)$\\
\refstepcounter{tbn}\label{pcsheVx8}\thetbn& \ref{pcsheV5} &
$x$ &  $M,\;$ $\p_t,\;$ $\p_x+tM,\;$ $G(2t)+t^2M,\;$ $D(2t)+G(3t^2)+t^3M$\\
\refstepcounter{tbn}\label{pcsheVx9}\thetbn& \ref{pcsheV5} &
$x^2+i\widehat\gamma$
&$M,\;$ $\p_t,\;$ $G(e^{2t}),\;$ $G(e^{-2t}),\;$ $D(e^{4t})$\\
\hline
\end{tabular}
\end{center}}

\vspace{0.5ex}

{\begin{center}
Table 3. Classification of the subclass $V=V(x)$ if $\gamma=4.$ \\
Here $\nu,\alpha,\beta\in\mathbb{R},$ $\nu\ne 0,$ $(\alpha,\beta)\ne(0,0).$
\\[1ex]\footnotesize
\renewcommand{\arraystretch}{1.3}
\setcounter{tbn}{0}
\begin{tabular}{|c|c|c|l|}
\hline\vspacebefore
N & N$_1\!\!$& $V$ & \hfill {Basis of $A^{\max}$\hfill} \\
\hline
\label{pcsheg4Vx0}\thetbn\vspacebefore& \ref{pcsheV6} &
$V(x)$ &  $M,\;$ $\p_t$ \\
\hline
\refstepcounter{tbn}\label{pcsheg4Vx1}\thetbn& \ref{pcsheV7} &\vspacebefore
$(\alpha+i\beta)x^{-2}$ &  $M,\;$ $\p_t,\;$ $D(t),\;$ $D(t^2)$\\
\refstepcounter{tbn}\label{pcsheg4Vx2a}\thetbn& \ref{pcsheV7} &
$x^2+(\alpha+i\beta)x^{-2}$
&   $M,\;$ $\p_t,\;$ $D(e^{4t}),\;$ $D(e^{-4t})$\\
\refstepcounter{tbn}\label{pcsheg4Vx2b}\thetbn& \ref{pcsheV7} &
\hspace*{1mm}$-x^2+(\alpha+i\beta)x^{-2}$\hspace*{1mm}
&   $M,\;$ $\p_t,\;$ $D(\cos 4t),\;$ $D(\sin 4t)$\\
\refstepcounter{tbn}\label{pcsheg4Vx3}\thetbn& \ref{pcsheV4} &
$i$ &  $M,\;$ $\p_t,\;$ $\p_x,\;$ $G(t)$ \\
\refstepcounter{tbn}\label{pcsheg4Vx4}\thetbn& \ref{pcsheV4} &
$x+i\nu$ &  $M,\;$ $\p_t,\;$ $\p_x+tM,\;$ $G(2t)+t^2M$\\
\refstepcounter{tbn}\label{pcsheg4Vx5}\thetbn& \ref{pcsheV2} &
$-x^2+i\nu$ &  $M,\;$ $\p_t,\;$ $G(\sin2t),\;$ $G(\cos2t)$\\
\refstepcounter{tbn}\label{pcsheg4Vx6}\thetbn& \ref{pcsheV3} &
$x^2+i\nu$ &  $M,\;$ $\p_t,\;$ $G(e^{2t}),\;$ $G(e^{-2t})$\\
\refstepcounter{tbn}\label{pcsheg4Vx7}\thetbn& \ref{pcsheV5} &
0 &  $M,\;$ $\p_t,\;$ $\p_x,\;$ $G(t),\;$ $D(t),\;$ $D(t^2)$\\
\refstepcounter{tbn}\label{pcsheg4Vx8}\thetbn& \ref{pcsheV5} &
$x$ &  $M,\;$ $\p_t,\;$ $\p_x+tM,\;$ $G(2t)+t^2M,$ \\
& & & $D(2t)+G(3t^2)+t^3M,\;$ $D(4t^2)+G(4t^3)+t^4M$\\
\refstepcounter{tbn}\label{pcsheg4Vx9a}\thetbn& \ref{pcsheV5} &
$x^2$ & $M,\;$ $\p_t,\;$ $G(e^{2t}),\;$ $G(e^{-2t}),\;$ $D(e^{4t}),\;$ $D(e^{-4t})$\\
\refstepcounter{tbn}\label{pcsheg4Vx9b}\thetbn& \ref{pcsheV5} &
$-x^2$ & $M,\;$ $\p_t,\;$ $G(\cos 2t),\;$ $G(\sin 2t),\;$ $D(\cos 4t),\;$ $D(\sin 4t)$ \hspace*{8mm} \\
\hline
\end{tabular}
\end{center}}

\noindent {\it Proof:} Let $V=V(x)$ and
$A^{\max}(V)\ne A^{\mathrm {ker}}_{V_t=0}.$
Consider an arbitrary operator $Q=D(\xi)+G(\chi)+\lambda M\in A^{\max}(V).$
Under Lemma's assumption, the condition~(\ref{classcondforcshewp}) implies
a set of equations on $V$ of the general form
\[
(ax+b)V_x+2aV=c_2x^2+c_1x+\tilde c_0+ic_0, \qquad \textrm{where}\qquad
a,b,c_2,c_1,\tilde c_0,c_0=\mathrm{const}\in\mathbb{R}.
\]
The exact number $k$ of such equations with the linear independent sets of coefficients
can be equal to either 1 or 2.
(The value $k=0$ corresponds to the general case $V_t=0$
without any extensions of $A^{\max}.$)

For $k=1$ $(a,b)\ne(0,0)$ and there exist two possibilities $a=0$ and $a\ne 0.$
If $a=0$ without loss of generality we can put $b=1$.
Then condition~(\ref{classcondforcshewp}) results in $\xi_t=0,$ $c_2=c_0=0,$ i.e.
$V_x=c_1x+\tilde c_0,$ and then $k=2$ that it is impossible.

Therefore, $a\ne 0$ and we can put $a=1.$
$\tilde c_0,b=0\!\!\mod G^{\Equiv}_{V_t=0}\!.\,$
Condition~(\ref{classcondforcshewp}) results in
$\chi=0$ (then $c_1=0$), $\lambda_t=0,$ $\widehat\gamma\xi_{tt}=2c_0\xi_t$
and $\widehat\gamma c_2=c_0^2.$
For $\gamma=4$ $c_0=0$ and $c_2\in\{-4,0,4\}\!\!\mod G^{\Equiv}_{V_t=0}$ and
these possibilities in the value of $c_2$ give Cases~3.\ref{pcsheg4Vx1}--3.\ref{pcsheg4Vx2b}.
If $\gamma\ne 4$  we obtain Cases~2.\ref{pcsheVx1} ($c_0=0$)
and~2.\ref{pcsheVx2} ($c_0\ne 0$).

The condition~$k=2$ results in $V=d_2x^2+d_1x+\tilde d_0+id_0.\,$
$\tilde d_0=0\!\!\mod G^{\Equiv}_{V_t=0}\!.\,$
Considering different possibilities for values of the constants $d_2,$ $d_1$ and $d_0$
and taking into account the value of $\gamma$ (either $\gamma\ne4$ or $\gamma=4$),
we obtain all the other classification cases:
\begin{gather*}
d_2=d_1=d_0=0\;\to\;2.\ref{pcsheVx7}, 3.\ref{pcsheg4Vx7}; \qquad
d_2=d_1=0,\;d_0\ne 0\;\to\;2.\ref{pcsheVx3}, 3.\ref{pcsheg4Vx3}; \\[1.2ex]
d_2=d_0=0,\;d_1\ne 0\;\to\;2.\ref{pcsheVx8}, 3.\ref{pcsheg4Vx8}; \qquad
d_2=0,\;d_0,d_1\ne 0\;\to\;2.\ref{pcsheVx4}, 3.\ref{pcsheg4Vx4}; \\[1.2ex]
d_2<0,\; (d_0,\widehat\gamma)\ne(0,0)\;\to\;2.\ref{pcsheVx5}, 3.\ref{pcsheg4Vx5};\qquad
d_2<0,\; d_0=\widehat\gamma=0\;\to\;3.\ref{pcsheg4Vx9b};\\[1.2ex]
d_2>0,\;\widehat\gamma^2d_2\ne d_0^2\;\to\;2.\ref{pcsheVx6}, 3.\ref{pcsheg4Vx6};\qquad
d_2>0,\;\widehat\gamma^2d_2=d_0^2\;\to\;2.\ref{pcsheVx9}, 3.\ref{pcsheg4Vx9a}.
\end{gather*}

\begin{note}
To prove Theorem~\ref{theorem.gc.pcshe}, it is sufficient to consider
only the case $k=1,$ $a\ne0$ in Lemma~\ref{classification.cshewp.Vt.0}
since other cases of extensions
of~$A^{\max}(V)$ with $V=V(x)$ admit operators of the form $G(\chi)+\lambda M$
($\chi\ne 0$) and, therefore (by Corollary~\ref{criteria.equiv.cshewp}),
are equivalent to Cases~\mbox{1.1--1.5}.
\end{note}

\begin{note}\label{ExplicitFormForTransVtVx}
The number $N_1$ for each line of Tables~2 and~3 is equal to the number of
the same or equivalent case in Table~1. The corresponding equivalence
transformations have the form~(\ref{eqtranspcshe})
where the functions $T,$ $X$ and $\Psi$ are as follows:
\begin{gather*}
2.\ref{pcsheVx2},\; 3.\ref{pcsheg4Vx2a} \to1.\ref{pcsheV7},\;
2.\ref{pcsheVx6},\; 3.\ref{pcsheg4Vx6} \to1.\ref{pcsheV3}
\left(\tilde\nu=\dfrac{\widehat\gamma-\nu}4\right), \;
2.\ref{pcsheVx9},\; 3.\ref{pcsheg4Vx9a} \to1.\ref{pcsheV5}{:}\quad
T=-e^{-4t},\;X=\Psi=0;
\\[1.4ex]
3.3 \to1.7,\quad   
2.\ref{pcsheVx5},\; 3.\ref{pcsheg4Vx5} \to1.\ref{pcsheV2}(\tilde\nu=\nu), \quad
3.\ref{pcsheg4Vx9b} \to1.\ref{pcsheV5}{:}\quad T=\tan 2t,\;X=\Psi=0;
\\[1.4ex]
2.\ref{pcsheVx4},\; 3.\ref{pcsheg4Vx4}\to 1.\ref{pcsheV4}{:}\quad
T=|\nu|t,\;X=-\sqrt{|\nu|}\,t^2,\;\Psi=\dfrac{t^3}3;
\\[1.4ex]
2.\ref{pcsheVx8},\; 3.\ref{pcsheg4Vx8}\to 1.\ref{pcsheV5}{:}\quad
T=t,\;X=-t^2,\;\Psi=\dfrac{t^3}3.
\qquad
\end{gather*}
\end{note}

\vspace{0.5ex}
Remark~\ref{ExplicitFormForTransVtVx} completes the proof of Theorem~\ref{theorem.gc.pcshe}.

\section{Conclusion}

The results of group classification obtained in this paper can be extended to a
more general class of $(1+n)$-dimensional NSchEs with potentials
\begin{equation}\label{generalNSchEwithPotential}
i\psi_t+\Delta\psi+F(\psi,\psi^*)+V(t,\vec x)\psi=0,
\end{equation}
where the $F=F(\psi,\psi^*)$ is an arbitrary complex-valued
smooth function of the variables~$\psi$ and~$\psi^*$.
We have already described all possible inequivalent forms of the
parameter-function~$F$ (without any restriction on the dimension $n$)
for which an equation of the form~(\ref{generalNSchEwithPotential})
with a some potential $V$ has an extension of the maximal Lie invariance
algebra.
We believe that the classification method suggested in this paper
can be effectively applied to complete the group classification
in~(\ref{generalNSchEwithPotential}) for the small values of $n.$
This method can be also a tool to investigate symmetries of other classes of PDEs,
and we will attempt to prove general statements on its applicability.

Another direction for our future research to develop the above results is
construction of both invariant and partially invariant exact solutions
of equations having the form~(\ref{schr}) by means of using found Lie symmetries,
and knowledge of explicit forms for equivalence transformations (see Theorem~\ref{NSchEwithPNPTheoremOnGequiv} and
Remark~\ref{ExplicitFormForTransVtVx}) allows us to reduced consideration of known
stationary potentials to simpler $x$-free ones.
We also plan to study conditional and generalized symmetries of~(\ref{schr})
to find non-Lie exact solutions.

As it was shown by Carles~\cite{Carles2002lanl}, the equivalence transformations~(\ref{eqtranspcshe})
also give an easy and effective way to produce new results on existence, uniqueness, estimations,
etc. of solutions for some equations~(\ref{schr}) by means of using known results on other potentials.

\subsection*{Acknowledgements}

The authors are grateful to
Profs. V.~Boyko, A.~Nikitin, I.~Yehorchenko and A.~Zhalij
for useful discussions and interesting comments.
The research of NMI was supported by National Academy of Science of Ukraine
in the form of the grant for young scientists.
ROP appreciate Prof.~F.~Ardalan
(School of Physics, Institute for Studies in Theoretical Physics and Mathematics, Tehran)
for hospitality and support during writing this paper.

\end{document}